# Nanoscale spatial resolution probes for Scanning Thermal Microscopy of solid state materials

P. Tovee[1*], M. Pumarol[1], D. Zeze[2], Kevin Kjoller[3], and O. Kolosov[1].

[1]Physics Department, Lancaster University, Lancaster, UK.

[2]School of Engineering & Computer Sciences Durham University, Durham, UK

[3]Anasys Instruments, Santa Barbara, CA, USA

*corresponding author, p.tovee@lancaster.ac.uk

## Abstract

Scanning Thermal Microscopy (SThM) uses micromachined thermal sensors integrated in a force sensing cantilever with a nanoscale tip can be highly useful for exploration of thermal management of nanoscale semiconductor devices. As well as mapping of surface and subsurface properties of related materials. Whereas SThM is capable to image externally generated heat with nanoscale resolution, its ability to map and measure thermal conductivity of materials has been mainly limited to polymers or similar materials possessing low thermal conductivity in the range from 0.1 to 1 $Wm^{-1}K^{-1}$, with lateral resolution on the order of 1 μm. In this paper we use linked experimental and theoretical approaches to analyse thermal performance and sensitivity of the micromachined SThM probes in order to expand their applicability to a broader range of nanostructures from polymers to semiconductors and metals. We develop physical models of interlinked thermal and electrical phenomena in these probes and their interaction with the sample on the mesoscopic length scale of few tens of nm and then validate these models using experimental measurements of the real probes, which provided the basis for analysing SThM performance in exploration of nanostructures. Our study then highlights critical features of these probes, namely, the geometrical location of the thermal sensor with respect to the probe apex, thermal conductance of the probe to the support base, heat conduction to the surrounding gas, and the thermal conductivity of tip material adjacent to the apex. It is furthermore allows us to propose a novel design of the SThM probe that incorporates a multiwall carbon nanotube (CNT) or similar high thermal conductivity graphene sheet material with longitudinal dimensions on micrometre length scale positioned near the probe apex that can provide contact areas with the sample on the order of few tens of nm. The new sensor is predicted to provide greatly improved spatial resolution to thermal properties of nanostructures, as well as to expand the sensitivity of the SThM probe to materials with heat conductivity values up to 100-1000 $Wm^{-1}K^{-1}$.

## I. Introduction

Scanning thermal microscopy is a well-established tool for investigating nanostructures that has the ability to provide sub-µm lateral resolution.[1-6] SThM's working principle is based on the scanning of a thermal sensor with a sharp thermally conductive tip (often also used as a heater) across a sample surface. The tip is mounted on a force sensitive cantilever and the feedback loop is used in order to maintain a constant tip-surface force while scanning across the sample surface in a raster way – an approach well known since the invention of the atomic force microscope (AFM).[7] With the tip in contact with the sample, that has a different temperature, the heat transfer between the tip and the surface changes the sensor temperature. In principle, SThM should be an ideal tool for exploring heat transfer in modern semiconductor or nano-electronic devices and imaging their surface and subsurface features, owing to its intrinsic sensitivity to local material properties and the ability of thermal waves to propagate in the material. Unfortunately, SThM applicability for mapping local thermal properties of materials has been mostly limited to imaging of polymeric materials with thermal conductivity, $k$, vales between 0.1 to a few $Wm^{-1}K^{-1}$ with a quite limited ability to differentiate between thermal properties of materials of higher thermal conductivity.[2,8,9] Moreover, in these cases, the lateral resolution reported in SThM is mainly reported on the µm scale[3]. As most semiconductors, optoelectronic materials and metals have thermal conductivities between 40 to 500 $Wm^{-1}K^{-1}$, and characteristic feature sizes in modern devices are on the order of few tens of nm, it is essential to find out an approach to expand SThM sensitivity to such materials and to improve SThM's lateral resolution. In essence, these are two goals we are addressing in this paper using a coupled theoretical modelling and experimental approach.

While the original SThM probes were made of Wollaston wire, where the 5 µm thick Pt wire served both as a heater and a resistive thermal sensor providing micrometre lateral resolution,[2] modern high resolution thermal probes are fabricated using batch micromachining processes[10,11] using materials such as silicon, silicon oxide or silicon nitride.[12] These probes have a tip apex in the few tens of µm range and, potentially, superior spatial resolution, but so far their reported thermal performance is far from perfect, as they are generally not sensitive to materials with high thermal conductivity. The reason is, as we show later in this paper, the weak thermal coupling between the thermal sensing element and the sample and the low thermal conductivity of the very apex of the probe. In order to improve thermal contact between sensor and sample, one could consider materials with higher thermal conductivity such as Carbon Nanotubes (CNT), the highest *k*-value available in nature, or other nanostructures incorporating a graphene sheet shown to have an extreme thermal conductivity of few thousands $Wm^{-1}K^{-1}$ *e.g.* folded graphene or nanocones.[13] Since CNT's discovery in 1991 by Iijima[14] a large amount of research effort has gone into studying their unique properties, particularly mechanical and electronic, with significantly smaller volume of research into local thermal properties of CNT's and no reports on them as thermal probes for scanning probe microscopy (SPM). As CNT's have already brought significant potential to SPM as ultimate resolution topography probes in tapping mode,[15] their extreme heat conductance coupled to their outstanding mechanical properties, suggests high potential of their application in SThM. A fully functional CNT thermal probe that could be reliably produced might greatly increase applicability and the thermal resolution of current SThMs.

Attaching of CNT probes to the tip can present certain challenges with current methods, such as picking up CNT's while scanning a SPM tip across a CNT array, followed by welding the CNT to the tip with a scanning electron microscope,[16] being relatively slow. Nevertheless, recent methods of in-situ grown CNT on the probes[15] can provide a good alternative solution. Other challenges to resolve are the mechanical stability of the tip in contact with the sample, heat transfer from the sample to the tip and the influence of ambient air.[17-19] With limited amounts of available experimental exploration, it is very important to create realistic quantitative physical models for such probes in order to guide the technological development and to test new design ideas. A good model will help with understanding the contributions of these difficulties and in building a better SThM probe design. Previous modelling of heat transfer between a SThM probe and a surface used an approximated analytical solution considering only the tip apex.[10] These models were focused on applications such as data storage, since sub-100 nm spatial resolution can be achieved.[20] In the limit of very small size of nanoscale contacts $L$ on the order of few nm in size and highly crystalline materials, with mean-free-path (MFP) for acoustic phonons $\Lambda_{ph}$ much larger than contact dimension, a ballistic regime of heat transfer becomes dominant. The theoretical analysis predicts then much higher thermal resistance[21,22] for Knudsen number $Kn = \Lambda/L >> 1$. In this paper we are considering mesoscopic dimensions of the contact of ~50 nm and features of the probe on the order of 1-10 μm that are higher than MFP in most elements of our nano-thermal probes, taking into the account their particular high aspect geometry. We address the effects of mesoscopic size in the nanoscale thermal probe below, based on the analysis of the transition regime between ballistic and diffusive thermal transport as detailed elsewhere.[23-25]

While SThM was used to study CNT's thermal conductivity,[6,26] few groups have experimental or modelling exploration of these systems.

In this paper, we analyse two major modalities of state-of-the-art microfabricated SThM probes, build realistic models of these probes, and then experimentally validate the models with experimental data. Based on these models, we then analyse the performance of each probe in application to a range of materials from very low to very high thermal conductivity (from polymers to semiconductors, metals and graphene sheet materials), in air and vacuum environments, thereby identifying key factors determining SThM sensitivity. Based on this study, we then propose a novel design of the SThM probe that is modified using a CNT or related high thermal conductivity graphene sheet material, and use our experimentally validated model to predict performance of such probe for nanoscale thermal measurements.

## II. Materials and methods
### A. Multi-scale modelling of nanoscale thermal probes

The two widely used types of micromachined SThM probes, namely, i) a dopped Si probe (DS), and a silicon oxide (or silicon nitride) ii) with a Pd thin film resistor probe (SP), have significant differences both in the material used, and probe geometry. The first one, DS, has geometry similar to a standard micromachined Si AFM probe (Fig. 1 (a)). It is made of single crystal Si that has high thermal conductivity ($k_{Si}$ ~130 Wm$^{-1}$K$^{-1}$), and a small radius of curvature of the tip of approximately 10 nm (therefore providing good spatial resolution in topography AFM mode). The moderately doped resistive temperature sensing part of this probe is separated from the actual contact with the sample by the end section of the cantilever and the length of the probe tip. DS probes are often used for heating polymeric samples to study their thermal transitions.[3] The second type, SP probe, is made of low thermal conductivity silicon oxide ($k$ ~1 Wm$^{-1}$K$^{-1}$), has a Pd resistive temperature sensor that is positioned at the very end of a flat beak-like tip at the end of the cantilever (Fig. 2 (a)). Radius of curvature of SP tip is on the order of 50 nm and although it is not as sharp as of DS probe, it is generally considered as of better sensitivity to the sample thermal properties and it is more often used to produce thermal conductivity images. The ideal probe would have the spatial resolution of the DS probe, and the thermal sensitivity of the SP sensor.

Given complex 3D geometry of these probes, heterogeneity of the probe materials, thermal effects of surrounding media and the sample, and distributed heat generation and sensing element, the most appropriate method to describes functioning of such probes should include 3D modelling of all thermal and electrical phenomena. The electrical transport given scale of the heating elements is on the order from 500-1000 nm and their thickness of 100-1000 nm can be adequately described by the Ohm's law.[27] The thermal transport in most elements of these probes (with characteristic dimensions on the order of ~µm) given MFP for heat transport by phonons in insulating parts of the probe (and corresponding length for electrons in metallic parts of the probe) ranging from low 10s of nm in oxides and nitrides to ~100 nm in single crystalline Si, can be described by diffusive heat transfer equation

$$\rho C_P \frac{\partial T}{\partial t} - \nabla(k\nabla T) = Q \qquad (1)$$

where $\rho$ is the density of material, $C_P$ is the heat capacity, $k$ the thermal conductivity and $Q$ the heat source.

The main area where deviations from the diffusive heat transport should be considered is the a) very end of the probe tip and b) probe-surface contact for high thermal conductivity samples with large MFP (*i.e.* non-metals) where ballistic heat transport can be significant.[22-24] The SiO$_2$ end of SP probe have the effective MFP on the order of $\Lambda_{SiO2} \approx$ 10 nm and 50 nm probe end and therefore Knudsen number for heat transfer $Kn \approx$ 0.2. As described in,[25] the thermal conductivity in the transition diffusion-ballistic regime $k_{db}$ can be with good approximation estimated as the decrease of the effective heat conductivity $k$, with

$$k_{db} = \frac{k}{2(\pi Kn)^2}\left[\sqrt{1 + (2\pi Kn)^2} - 1\right] \qquad (2)$$

and corresponding increase in total thermal resistance of tip-surface contact $R_c$ [22] to

$$R_c = \frac{1}{2kL}\left[1 + \frac{8}{3\pi}\text{Kn}\right] \quad (3)$$

Low Knudsen numbers $Kn \ll 1$ in (2) correspond to the diffusive transport with $k_{db} \approx k$ and high $Kn \gg 1$ correspond to ballistic transport with $k_{db} \approx k/(\pi\ Kn)$. Clearly, SP probe with Knudsen number $Kn \approx 0.2$ is well described by the diffusive heat transfer. The Si probe end of DS probe have the characteristic cross-section dimension of $L_1 \sim 50$ nm (with low angle conical broadening of approximately $L_2 \approx 500$ nm length) that leads to Knudsen number for this probe of $Kn \approx 4\text{-}5$ accounting for the MFP of $\Lambda_{Si} \sim 250$ nm,[22] increase in ballistic component of heat resistance $R_{cb}$ to $R_{cb} \approx 20/(3\pi k L_1) \approx 2/(kL_1)$. At the same time, due to the elongated geometry of DS probe tip, the diffusion component of thermal resistance of SP probe tip end $R_{cd}$ obeying Fourier law will also be increased by the $L_2/L_1$ factor with $R_{cd} = L_2/(k\pi L_1^2) \approx 3/(kL_1)$ making it similar to the ballistic resistance $R_{cb}$, indicating that diffusion heat transfer is a reasonable approximation for DS probe as well.

Clearly, for insulating single crystalline materials of very high thermal conductivity, such as diamond, Si, sapphire and graphene layers the heat transport on the material side is ballistic in the vicinity of the ~50 nm contact area, and in this case, the appropriate consideration of modification of thermal resistance as per[22,24,25] using equations (2) and (3) and our SThM measurements will generally provide lower values of thermal conductivity with the factor of $1/(\pi\ Kn)$.

In order to analyse the detail performance of these probes, we first used finite elements analysis (FEA) approach based on COMSOL Multiphysics® to create a realistic three dimensional (3D) model of thermal transport in the probes.[28] This included joule heating of the probe and a temperature dependent resistance of the thermal resistive sensor, exploring ability of such sensors to both generate and evaluate thermal flow to the sample. We used "AC/DC", "Thermal" and "MEMS" modules of COMSOL®, both in static and time dependent solver configurations. All models were created and debugged in vacuum environment, with air environment subsequently introduced as a block of air enclosing the entire cantilever and the sample.

The example of our simulation results of the cantilever and sample can be seen in figure 1 (b). An electrical joule heating model was fully coupled to the thermal one. The initial conditions for the model were that all subdomains were set to a room temperature of 293 K. The thermal boundary conditions for the outer surfaces of the sample and the air block were thermally anchored at 293 K, whereas all inner boundaries were presumed to have continuity of temperature, as in the first approximation, thermal resistances at the boundaries can be neglected. Electrical boundary conditions were a set potential and a ground applied to the two probe leads, whereas the resulting current (and, therefore, resistance of the probe representing the measured in SThM probe temperature) was one of the results of the simulation. The mesh used was coarse for larger areas but finer for the narrow constrictions near the tip. Making the mesh an order of magnitude smaller finer than the one we generally used, lead only to minor changes of the temperature difference on the order of 0.3%, suggesting the adequate quality of the mesh. Another significant output of the simulation was a 3D distribution of the temperature in the probe, surrounding air (if present), and in the sample. We have used the generic data produced by the probe manufacturers as a starting point for the model, with SEM characterization of geometry of the individual probes and tune-up of the model to further match experimental measurements of the probes. The probe - sample contact was represented as a circle area with 50 nm diameter, but these and other parameters of simulations could be varied as we will mention later in the discussion chapter.

Overall dimensions of the block of air surrounding the tip were on the order of sub millimetre range, making it possible to consider only heat conductance and neglecting convection contribution. Also, whereas nanoscale heat transfer in gases is known to have a ballistic character at distances below 50 nm,[17] and as the smallest dimensions in our model were on the order of 50 nm or larger, it was not essential to take these effects into account. This model was then expanded to study dynamic heating, which, although being outside of the scope of this paper, allowed us to further confirm the validity of the physical models as well as provide the basis for studying dynamic SThM, *e.g.* SThM in tapping mode.

The DS probe was invented by King, *et al.*[29] and marketed by Anasys Instruments (ThermaLever probe AN2-300). An electron microscope (SEM) picture of the DS probe can be seen in figure 1 (a) with its typical electrical and thermal parameters given in the caption.

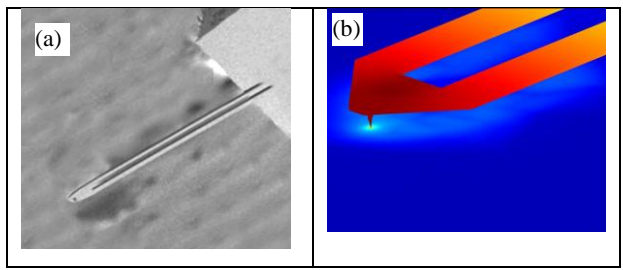

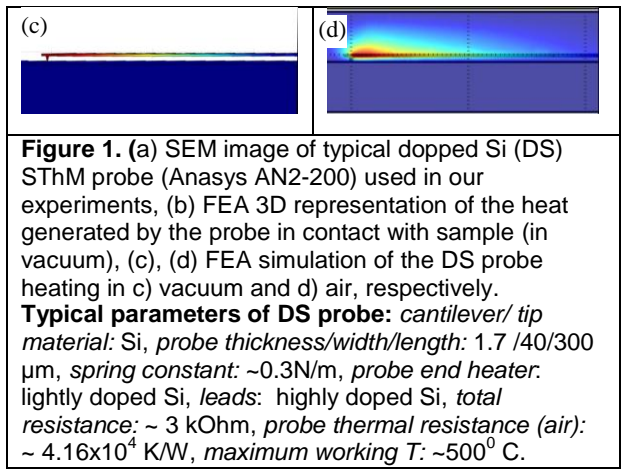

**Figure 1. (**a) SEM image of typical dopped Si (DS) SThM probe (Anasys AN2-200) used in our experiments, (b) FEA 3D representation of the heat generated by the probe in contact with sample (in vacuum), (c), (d) FEA simulation of the DS probe heating in c) vacuum and d) air, respectively.
**Typical parameters of DS probe:** *cantilever/ tip material:* Si, *probe thickness/width/length:* 1.7 /40/300 μm, *spring constant:* ~0.3N/m, *probe end heater*: lightly doped Si, *leads*: highly doped Si, *total resistance:* ~ 3 kOhm, *probe thermal resistance (air):* ~ 4.16x10$^4$ K/W, *maximum working T:* ~500$^0$ C.

The probe is made by standard Si micromachining process and has two low resistance highly doped legs of 300 μm length joined by the high resistance heating element situated near the tip which is about 5 μm high. The heating element creates the joule heating when an electric voltage is applied across the bases of the two legs.

The SP probe is made of silicon oxide, $SiO_2$, (or, more recently, of silicon nitride, $Si_3N_4$) with a palladium resistor (SP probe) as reported, for example, by Weaver, *et al.*[30] and is currently marketed by Kelvin Nanotechnology. The tip radius of SP probe is about 50 nm with two gold strips carrying the current down to the tip where a thin higher resistance palladium film generates the heat. With the thermal element (that is both a heater and a temperature sensor) positioned close to the tip apex, and therefore close to the sample, one expects the thermal sensitivity to increase. A SEM image of a SP sensor can be seen in figure 2 along with the typical electrical and thermal parameters for the probe given in the caption. As can be seen, the materials and geometry of the probe is quite different from the DS one. Both probes had 100 Ω resistors built into each leg and Au contacts. The properties of the probe were adjusted to account for these resistors.

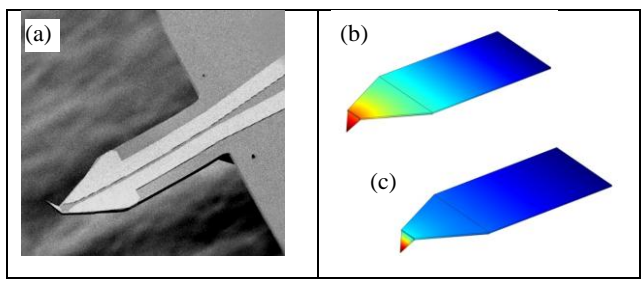

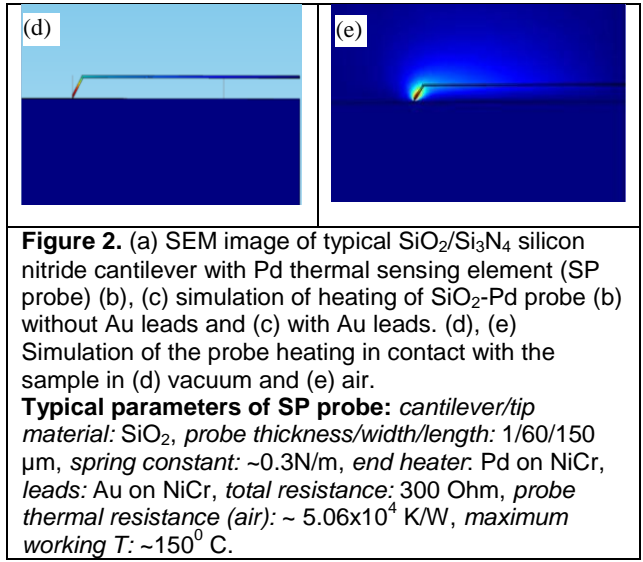

**Figure 2.** (a) SEM image of typical SiO$_2$/Si$_3$N$_4$ silicon nitride cantilever with Pd thermal sensing element (SP probe) (b), (c) simulation of heating of SiO$_2$-Pd probe (b) without Au leads and (c) with Au leads. (d), (e) Simulation of the probe heating in contact with the sample in (d) vacuum and (e) air.
**Typical parameters of SP probe:** *cantilever/tip material:* SiO$_2$, *probe thickness/width/length:* 1/60/150 µm, *spring constant:* ~0.3N/m, *end heater:* Pd on NiCr, *leads:* Au on NiCr, *total resistance:* 300 Ohm, *probe thermal resistance (air):* ~ 5.06x10$^4$ K/W, *maximum working T:* ~150$^0$ C.

The modelling of SP probe with and without Au leads (figure 2 (b) and (c)) shows tremendous qualitative and quantitative difference due to much higher thermal conductivity of Au than SiO$_2$/Si$_3$N$_4$ cantilever, even though the thickness of Au is much smaller that the thickness of the cantilever (note that in the probe design with the Au layer corresponds to the more realistic probe model, and the temperature increase is limited mainly to the tilted beak of the sensor, Figure 2 (c)). The Au model has been shown to correspond much closer to our experimental measurements. In modelling, we have used Si and Poly(methyl methacrylate) (PMMA) as model samples for both probes as these have highly different thermal conductivities and could be easily prepared for the experimental measurements. The model was tested with different tip contact diameters and we observed that the larger area of the thermal contact with, *e.g.* Si sample, the higher is the heat flow to the sample and the lower the temperature of the tip, as it would be expected, both in vacuum and in air. Although the contact area could not be arbitrarily changed experimentally to be directly compared with the model, such calculations would allow an estimation of the actual contact area of the probe by comparing experimental results with the model.

## B. Experimental setup for SThM probe testing

In order to compare modelling results with the experimental data, the following experimental procedure was used. First, the thermal sensor was calibrated by linking it's resistance to sensor ambient temperature (using small applied voltage so that self-heating is negligible), then the increasing voltage was applied to the sensor and the temperature of Joule self-heating of the sensor was measured as a function of applied power. Finally, the tip of a SThM sensor was brought in contact and out of contact with the sample (in air and in high vacuum environments), while measuring its temperature. That allowed us to quantify heat transfer from the sensor to the sample and to air. An additional experimental setup allowed dynamic measurements of the transient response of SThM probes and to compare it to the simulation results, which allowed further validation quantitative of the sensor model.

The experimental setup was based on high vacuum multifunctional SPM (HV NT-MDT Solver HV-AFM) encased in a dedicated chamber that can be either evacuated to approximately $10^{-7}$ torr or used at an ambient air pressure. This chamber was equipped with instrumental feedthroughs for the thermal probe. The SPM was suspended on springs with efficient eddy-current magnetic dampers; during vacuum operation the turbo-molecular pump with an oil-free scroll backing pump provided initial high vacuum, and the vibrations–free ion pump securing necessary vacuum during thermal measurements. The SPM system used a laser beam deflection system in order to measure forces acting on the SThM in contact mode, and allowed us to position the sensor in contact and out of contact with the test sample, as well as to monitor the force acting between the sample and the probe tip. The electrical measurement setup is shown in figure 3 (a), where the sensor was either a part of a voltage divider in series with the fixed resistor, or a part of an AC-DC electric bridge. In both cases the voltage excitation was provided by the precision function generator (Keithley 3390 50 MHz arbitrary waveform generator). In voltage divider mode a dedicated multimeter (Agilent 34401A 6.5 digits precision) in the ratiometric mode allowed us to measure the resistance of the probe as a function of the voltage at the probe (and, therefore, its temperature due to Joule heating). The probe temperature due to self-heating could be raised in excess of 100 $^0$C above ambient temperature. In the AC-DC bridge configuration, the bridge was balanced at room temperature and absence of SPM laser illumination (we measured that SPM laser deflection monitoring system provided additional heating of approximately 7 $^0$C in air to 15 $^0$C in vacuum), using variable resistor and capacitor.

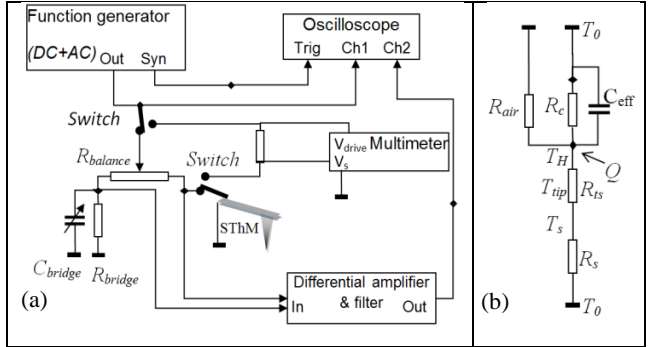

**Figure 3:** (a) Schematic diagram of AC and DC measurement SThM electronics. (b) Schematic diagram of equivalent heat resistances of the probe accounting for the conductance to the cantilever base ($R_c$), air ($R_{air}$), tip contact resistance to the sample ($R_{TS}$) and sample resistance ($R_S$). $C_{eff}$ represents non-zero time constant of the probe. Where $Q$ is the thermal power generated in the system, $T_H$ is the heater temperature and $T_0$ = 293 K.

The thermal calibration of the probes was performed using a temperature stabilized Peltier hot/cold plate (Torrey Pines Scientific, Echo Therm model IC20) at several temperatures from room temperature to 100 $^0$C. As this calibration took place outside the SPM, where there was no heating from the laser. The self-heating of the probe was measured using the same setup, and the results of these measurements are given in figure 4. It should be noted that during self-heating, the distribution of the temperature over the probe is inhomogeneous (the end is hotter than the base, see figure 1 (c)-(d)). As the temperature resistive element is extended over some distance, a temperature deduced from the probe resistance measurements is effectively averaged over the area of the sensitive element. With this effect, while not exceeding fraction of a per cent for the DS probe, is much more notable in the SP probe with temperature differences as large as 30%. Therefore for the SP probe during FEA we used simulated "hot-cold plate" uniform distribution of the temperature to find the sensitivity of the SP probe to the temperature, and appropriately, non-uniform distribution while self-heating, in contact and in air. That allowed us to make fully justified comparison of the experimental and the simulated data, which would be very difficult without FEA analysis of real probe geometry. The only caution is that for both experimental and FEA data the "thermal resistance" for self-heating SP probe is in fact an averaged thermal resistance over the whole heater area. Dynamic measurements of sensor heating [31,32] were performed using an AC-DC bridge by applying a square voltage pulse using a function generator and detecting a heating response from the bridge via a differential amplifier and a digital oscilloscope (figure 3 (a)). It should be noted that one has to account for the effect of the laser illumination on thermal measurements. Such illumination increases the tip temperature, and, therefore, its resistance, which was clearly seen in the SP probe measurements. However, at the same time, for the DS probe where carriers are excited by laser light in the semiconductor, the overall resistance was slightly decreasing. We have used reference measurements to account for these phenomena.

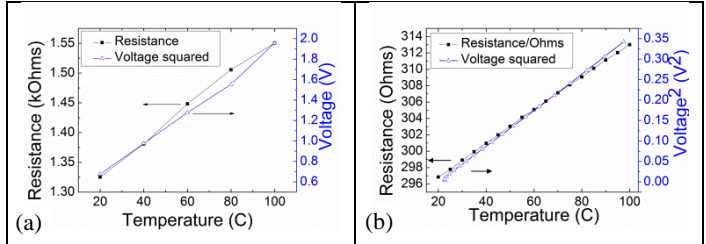

**Figure 4.** Calibration of the probes – (a) DS, (b) SP probe. Left vertical axis *vs* horisontal axis - resistance as a function of the probe temperature; note close to linear dependence of probe resistance *vs* probe temperature. Right vertical axis *vs* horisontal axis – self heating of the probe due to applied DC voltage to the probe, horisontal axis - temperature rise caused by the self heating; quadratic scaling of right axis illustrates that the probe temperature increase is linearly proportional to the Joule heating power (~$V^2$) applied to the probe.

The thermal probe can be represented by an equivalent thermal resistances diagram (Fig. 3 (b)), where the thermal resistances, $R_T$, were calculated using the following equation

$$R_T = \frac{\Delta T}{Q} \qquad (4)$$

where *ΔT* is the difference in temperature between the probe and the ambient surroundings and *Q* is the power flow. In the Fig 3 (b), the thermal resistance of the cantilever to the cantilever base ($R_c$), thermal resistance to air ($R_{air}$), and the thermal resistance of the tip contact to the sample ($R_{TS}$) in series with the sample thermal resistance ($R_S$) are presented similar to electrical resistances. In this diagram, *Q* is the Joule heat generated by the current flowing through the sensor, $T_H$ is the heater temperature, and $T_0$ is the ambient temperature (293 K). Using these models, the total heat resistance of the probe $R_p$ is

$$\frac{dQ}{T_H - T_0} = \frac{1}{R_p} = \frac{1}{R_c} + \frac{1}{R_{air}} + \frac{1}{(R_{ts} + R_s)}. \qquad (5)$$

If the probe is out-of-contact with the sample and is in vacuum, the air and the sample heat transfer terms vanish, and then $R_c$ is calculated. Correspondingly, $R_{air}$, $R_{ts}+R_s$, can be found by placing the cantilever in an air environment and in contact with the sample. It is essential to note that the tip-sample contact and the intrinsic sample resistances are combined in the sum of $R_{ts}+R_s$ and cannot be easily separated. As a result, if contact resistance is too high, *i.e.*

$R_{ts} \gg R_s$ (the case of the sample with high thermal conductivity) the SThM loses its sensitivity to the sample thermal properties. This and other phenomena are analysed quantitatively later in this paper.

Test samples for the experimental measurements were the same materials as used in the modelling approach – namely PMMA ($k_{PMMA}$ ~0.19 Wm$^{-1}$K$^{-1}$) and single crystalline Si [001] ($k_{Si}$ ~ 130 Wm$^{-1}$K$^{-1}$). Si wafer and PMMA plate were glued to an AFM sample holder and a suitable area clear of any scratches or visible marks was used. The thermal resistance of the sensor was measured when the tip was in contact with the surface, with its dependence on the contact force (although observed and a subject of a follow-up work) being negligible for the purpose of current study.

## III. Results and discussion
### A. Experimental verification of SThM probe models

The simulation allowed us to observe important phenomena of thermal interaction between the nanoscale SThM tip and the surface. The first qualitative observation is that in vacuum environment the sensors that are out-of-contact with the sample, have approximately the same temperature for the whole tip area (Fig. 5 (a) and 6 (a), the temperature is given in false `thermal' colours where white corresponds to highest temperature, black to lowest). The situation is quite different for the tip in contact with the high conductivity material. In this case, most of the temperature drop happens at the very end of the tip whereas the sample is barely heated (Figs. 5 (b) and 6 (b)). This suggests that the limiting factor in the heat transfer between SThM sensor and the sample for high conductivity sample material is the very end of the tip. That corresponds to high thermal resistance $R_{ts}$ in respect to $R_s$ in the equivalent thermal circuit for the SThM tip (Fig 3 (b)). Conversely, in case of low conductivity material – such as PMMA, the main temperature drop occurs in the bulk of the material itself ($R_{ts} \ll R_s$) and the temperature distribution in the tip is approximately as uniform as in the non-contact case. These observations qualitatively explain why current SThM sensors can differentiate well between low thermal conductivity materials (*e.g.* polymers) as it is the sample property that defines the heat flow from the tip. Whereas they are incapable of efficiently distinguishing between high conductivity materials (*e.g.* semiconductors) as the SThM tip limits the heat flow. This matches well with the experimental results obtained with these sensors.[2] Evidently, a far better design for a SThM probe for these materials would be to have the heater and sensor at the very tip (such as in the SP probe). That should help deliver more of the heat directly to the sample and have the high heat conductivity material at the end of the tip. A quantitative study of SThM performance confirming these qualitative analysis is presented later in this paper.

Although SThM sensors in this study are evaluated for their DC performance, our model and experimental setup was also used to study the dynamic heating effects, using a square pulse, and AC heating, and analysing the response time of the sensor. The time constants showed the same order of magnitude for both simulation and the experimental measurements. In particular, for DS Si probe they were 0.4 ms (in experiment) and 0.8 ms (in simulation), whereas for SP they were 0.3 ms experimentally and 0.2 ms for the model.

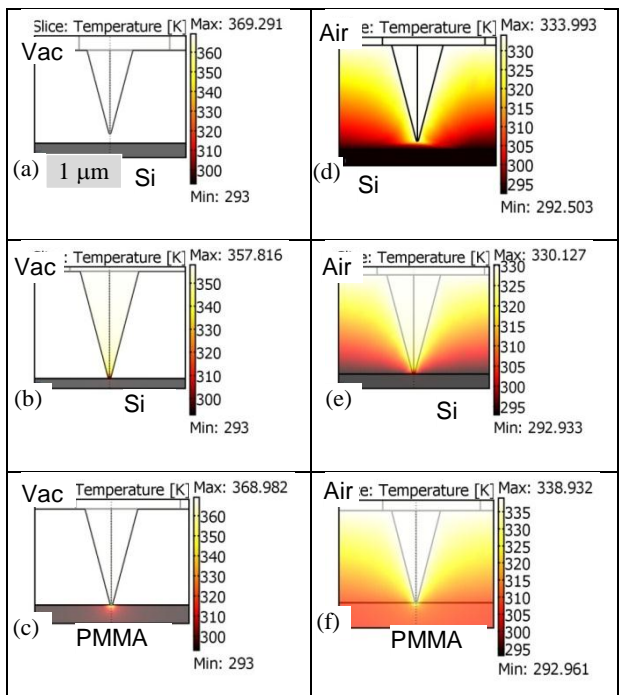

**Figure 5.** 3D simulation finite eliment modeling of of thermal response of DS probe in vacuum – (a), (b), (c) and in air – (d), (e), (f). Probe tip out-of-contact with the sample is shown in – (a), (d); in contact with Si - (b), (e) and PMMA (c), (f).
One can observe that probe out of contact with the sample (a) have the same temperature down to the end of the tip, whereas a probe in contact with a high heat conductive sample - Si (b) the main temperature drop occurs at the very end of the probe. Temperature increase of the sample with high thermal conductivity (Si, (b), (e)) is generally small and is negligible outside area of immediate contact, whereas it is fairly noticeable for the case of the lower heat conductivity material (PMMA, (c), (f)).

Quantitative analysis of the simulation data shows that moving from vacuum to air environment immediately decreases the maximum temperature attained by the sensor (fig 5 (d) and 6 (d)). The difference between the sensor temperature and ambient temperature (temperature of the sample and air far from the SThM sensor is assumed 293 K, or, approximately, 20 $^0$C, both in the simulation and the experiment) is reduced approximately by half (from 73 $^0$C to 43 $^0$C for the DS sensor and 73 $^0$C to 54 $^0$C for the DS sensor) when moving to air environment. The useful way for relating simulation and experimental data, is to compare the values for the total thermal resistances of the probe $R_p$. This data are given in the tables I (for DS probe) and II (for SP probe).

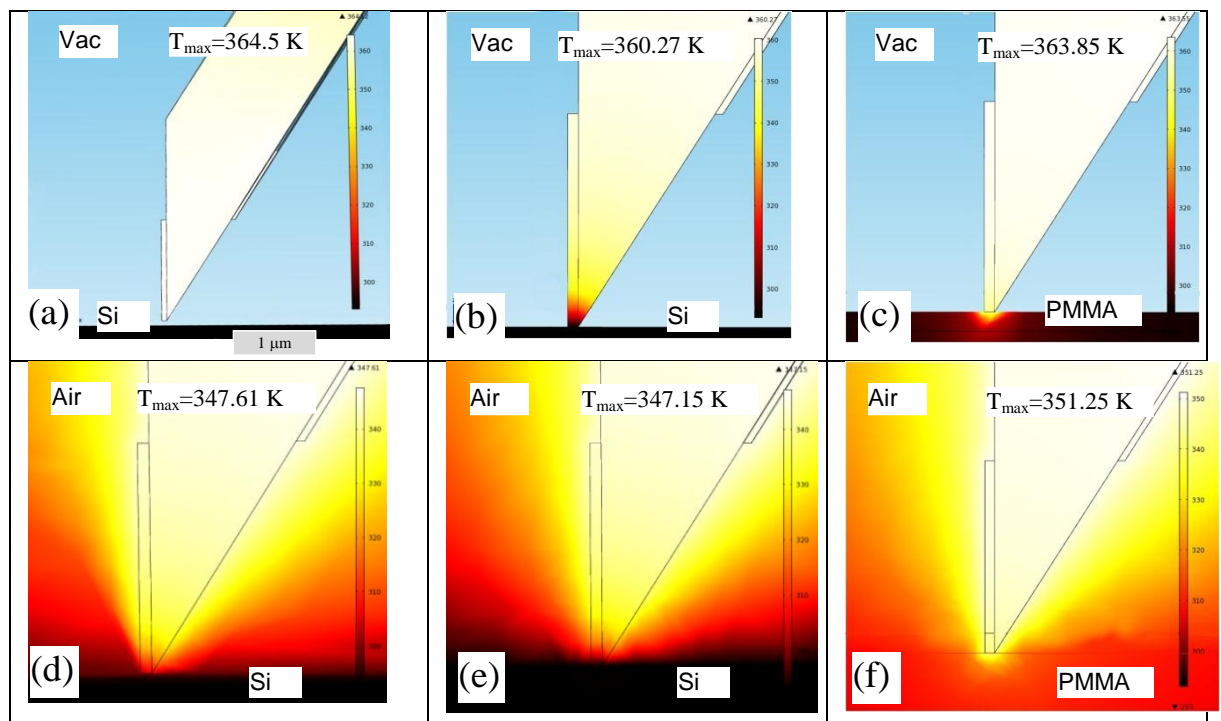

**Figure 6.** 3D simulation of thermal response of SP probe in vacuum – (a), (b), (c) and in air – (d), (e), (f). Probe tip out-of-contact with the sample is shown in – (a), (d); in contact with Si - (b), (e) and PMMA (c), (f). As for DS probe, the probe out of contact with the sample (a) have the same temperature down to the end of the tip, whereas for a probe in contact with a high heat conductivity sample, (b), the temperature drop occurs at the very end of the probe. The temperature increase of the sample with high thermal conductivity (Si, (b), (e)) is generaly small and is negligible outside the area of immediate contact, whereas it is fairly noticeable for the lower heat conductivity material (PMMA, (c), (f)).

In the tables shown below, further quantitative comparison of modelling and experimental results has been done by comparing sensor temperature change (and, corresponding thermal resistance change) when probe tip moves from out-of-contact position to contact with the sample. Corresponding changes in thermal resistance and ratios between contact and non-contact thermal resistances are given in tables I and II, with ratios between 2% and 10 % with these values consistently higher for the SP probe. This strengthens our assumption that the probe with the thermal sensor closer to the apex of the tip is more sensitive to the properties of the probed sample. There were inevitable deviations between the simulation and measurements, some reasons of these are i) the block of air we have used for the simulation was only 200 μm tall, as opposed to several millimetres in the experiment, ii) the exact geometry of the nanoscale tip-sample contact was unknown, and iii) our modelling approach considers all boundaries as continuity and does not take into account phonon reflection at boundaries (see note in the section B that confirms that this is a reasonable approximation for the system we consider). Also, iv) the water meniscus in air and heating from the AFM laser were not modelled. Whereas these do not influence main features of analysis and conclusions we can draw on the basis of these models, such features were the most likely explanation for the temperatures in the experiments reaching a few degrees less than in the model. The effect of the thickness of the air block was tested by doubling it's size from 200 μm to 400 μm that changed the temperature of the probe relative to the environment by less than 1%. This was negligible for the purpose of our calculations. Therefore it is fair to say that size of the air block is sufficient for our calculations.

It should be noted that for both the DS and SP probes the relative position of the sample and the cantilever while moving out of contact (for example by 50-100 nm) changes negligibly compared to the overall dimensions of the probe (10 to 100 μm). We assumed that the structure of heat flux (through the cantilever bases compared to through the air) should not change significantly. In order to prove this, we have integrated the fraction of heat flux going through the cantilever in and out of the contact in air environment, and have found that the changes in question were less than 0.3% confirming validity of our assumption.

**Table I.** Thermal resistances for the DS probe.

| Media | Contact/non-contact | Thermal resistance – experiment [K/W] | Thermal resistance – simulation [K/W] |
|---|---|---|---|
| Vacuum | n/contact | $5.33 \times 10^4$ | $4.72 \times 10^4$ |
|  | Contact w/Si | $5.18 \times 10^4$ | $4.62 \times 10^4$ |
|  | Ratio contact-n/contact | 1.028 | 1.022 |
| Air | n/contact | $4.16 \times 10^4$ | $2.34 \times 10^4$ |

**Table II.** Thermal resistances for the SP probe.

| Media | Contact/non-contact | Thermal resistance – experiment [K/W] | Thermal resistance – simulation [K/W] |
|---|---|---|---|
| Vacuum | n/contact | 8.0x10$^4$ | 5.35x10$^4$ |
| | Contact w/Si | 7.2x10$^4$ | 5.08x10$^4$ |
| | Ratio contact-n/contact | 1.11 | 1.053 |
| Air | n/contact | 5.06x10$^4$ | 3.39x10$^4$ |

Overall, the SThM probe parameters measured in our experiments are of the same order of magnitude as the FEA model data. With measured thermal resistances within the factor of 0.6 – 1.7 between simulations and experiments, and sufficiently close for qualitative analysis of the performance of the SThM probes. This meant we were able to use our models for the analysis of the performance of the existing probes and for the design of the new, superior performance nanoscale thermal probe.

These models assumed a temperature continuity across the tip-sample interface, effectively neglecting contact thermal resistance. It is known to be essential for thermal transport across dissimilar materials ("Kapitza resistance" across liquid He and metal[33]), and can be notable for other materials pairs[34]. For the SThM tip-sample pairs considered in this study where the phonon heat transport dominates, the interface thermal conductance (for immediate contact) is generally in the range between 80 and 300 MWm$^{-2}$K$^{-1}$. We explored the effects of such resistance in our FEA simulations to investigate it's effect on the sensitivity of the probe using vacuum models where such effect would be most noticeable. We then used a modified equation 5 to find the changes in tip-surface thermal resistance $R_{ts(i)}$ due to the presence of interfacial thermal resistance $R_i$, and to compare it with the tip-surface thermal resistance in the absence of $R_i$,

$$\frac{1}{R_P} = \frac{1}{R_c} + \frac{1}{R_{ts(i)}} \qquad (6)$$

In our analysis we considered the range of the sample thermal conductivities and contact thermal resistances values. As we expected, for the low to medium thermally conducting samples within the range of probe sensitivity (up to few tens Wm$^{-1}$K$^{-1}$ corresponding to polymers, oxides, nitrides, alloys and amorphous semiconductors) see Figure 7, accounting for interfacial thermal resistance provides only a small correction with $R_i$ at about 3.75x10$^6$ KW$^{-1}$ comparing to the total tip-sample thermal resistance $R_{ts(i)}$ of 1.49x10$^7$ KW$^{-1}$. That allows us to conclude that it is the temperature drop at the end of the SThM tip and at the constricted contact with the sample (fig. 5, 6) that is limiting the performance of SThM for probes we considered.

However for higher conducting samples like Si and the probe that is of higher thermal conductivity like CNT the results will be affected to a greater degree that is considered in the next section. Also, the contact imperfections with the solid-solid thermal contact existing across the part of the area of tip-surface contact (these are more likely in vacuum environment in the absence of water meniscus[35] and at low tip forces) may further contribute to the SThM thermal contrast, but these are beyond the scope of this paper.

### B. Analysis of SThM sensitivity to material thermal conductivity

A first major goal of scanning thermal microscopy is to identify materials by their thermal conductivity.[2,9] The measurable parameter in SThM is the 'temperature change of the heater' when the SThM tip is brought in-contact with the sample. The higher the thermal conductivity of the sample, the lower the temperature of the heater, the steeper such a dependence, and, hence, the better the sensitivity of the probe. The second important task of SThM is to create a local `hot spot' by locally heating a nanoscale area of the sample. This can be used, *e.g.* to probe the thermal transitions in the sample, as well as to explore functioning of the nanodevices.[3,17] The SThM's performance is then estimated by the temperature of *the sample* near the apex of the tip. Higher the temperature is the more efficient SThM is in affecting such local heating.

Using our validated models of the SThM probes, we have analysed performance of the DS and SP probes for these tasks with the results presented in Fig. 7. The left vertical axes in these graphs correspond to the temperature change of the heater as a function of the sample thermal conductivity whereas the right vertical axes correspond to the temperature of the sample near the apex of the tip. These results show that the optimal range of material sensitivity of both DS and SP SThM probes in vacuum (Figure 7 (a), (c)) is between 0.1 to few tens of $Wm^{-1}K^{-1}$. This generally covers a range of polymers, oxides, some III-V semiconductors, but stops short of the thermal conductivity of Si, Al, Cu that are widely used materials in the semiconductor and nanotechnology electronic industries. Also this range is well below the thermal conductivity of CNT and graphene materials. The SThM probe's ability to locally heat the sample generally follows the same trend (as seen in the graphs that correspond to the right vertical axes).

Placing these probes in air restricts their upper sensitivity range to even smaller values below 10 Wm$^{-1}$K$^{-1}$. The graphs show that a lower end of sensitivity apparently gets expanded well below 0.1 Wm$^{-1}$K$^{-1}$, with the total range of the temperature changes even exceeding those in vacuum. A careful analysis, though, shows that this expansion represents an effect of the heat conductance directly from the cantilever to the sample through the air layer. As this heat transfer happens on the length scale of the cantilever dimensions (20 - 100 μm) it does not carry any useful spatially resolved thermal information and is purely detrimental to the thermal resolution. As we discussed above in II.A, the range of thermal conductivities both SThM probes are sensitive to, is well within approximation of diffusive thermal transport.

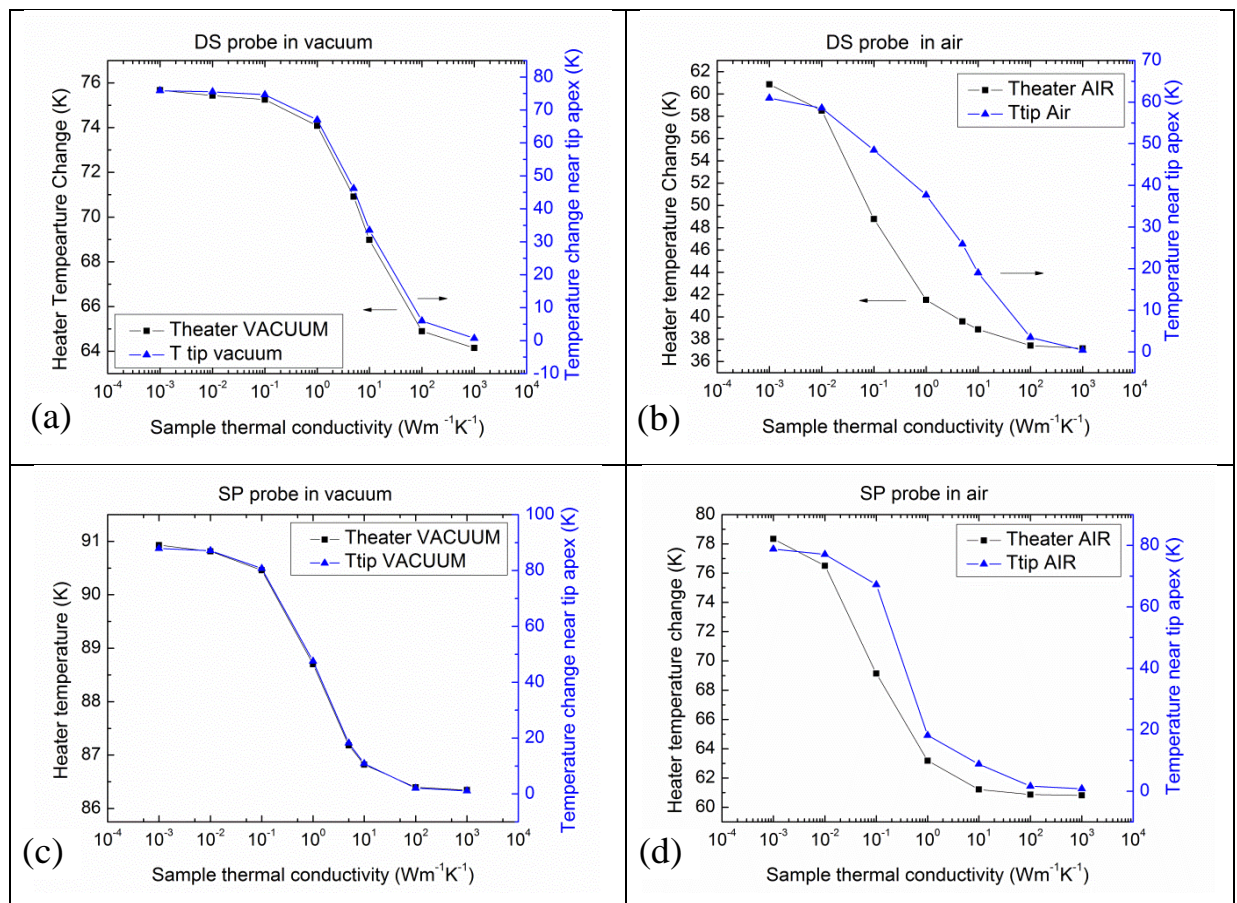

**Figure 7.** Sensitivity of SThM probes for materials of various thermal conductivity. DS probe ((a), (b)); SP probe - ((c), (d)); in vacuum - ((a), (c)); in air - ((b), (d)). Sample thermal conductivities range from 0.1 Wm$^{-1}$K$^{-1}$ (gases and porous materials) to 1000 Wm$^{-1}$K$^{-1}$ (graphite and graphene). Results show that for existing probes, an increase of sample temperature $T_t$ near the SThM tip apex is one order of magnitude smaller than the temperature of probe heater $T_h$ for any sample with thermal conductivity above 1 Wm$^{-1}$K$^{-1}$. At the same time, variations of the heater temperature $T_h$ (measure of the output signal of SThM) become practically insensitive to the sample properties for the samples with thermal conductivities above 10 Wm$^{-1}$K$^{-1}$.

Both DS and SP probes use either Si or silicon oxide. A material of higher thermal conductivity placed in the immediate contact between the tip and sample would be beneficial for their thermal sensitivity. An ideal probe would have high thermal sensitivity and a sharp tip. To achieve this, the heater must be close to the tip apex and the tip must be highly thermally conducting and reasonably sharp. As SP probes have the heater positioned in such a way (with currently a quite low thermal conductivity silicon oxide at the apex), we will explore what happens when a CNT is incorporated in the apex of the SP SThM probe. CNTs have very high thermal conductivity, good mechanical stability and high aspect ratio.

### C. A superior thermal probe – SThM-CNT probe.

Using our previous models, we have modified our SP probe model by adding a multiwall CNT at the end of the probe. Such CNT can be replaced by a sheet of a multilayer of graphene or a multiwall nanocone that will act as a heat conducting element in direct contact with the sample. Graphical representation of results of simulations for such device can be found in Figure 8.

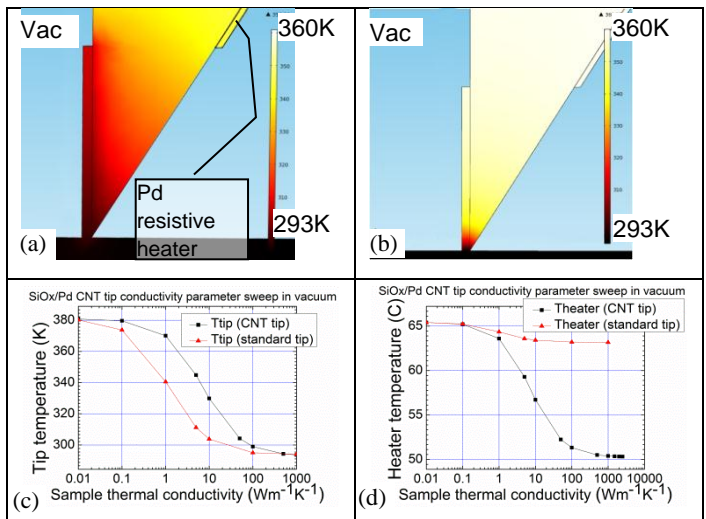

**Figure 8.** Results of simulation of thermal performance of CNT-SP SThM probe.
(a) Temperature distribution near the apex of CNT modified SP probe in contact with Si sample in vacuum. The tip and heater area have much more effective thermal link with the sample compared to existing $SiO_2$/Pd probe in vaccum. That results in a notable increase of the local sample temperatures under the SThM tip (b) compared with standard $SiO_2$ tip. (c) for high conductivity samples of 10-100 $Wm^{-1}K^{-1}$ as well ~ 10 fold increase of the SThM sensitivity to the thermal conductivity of such materials. A greater range of temperature drops in the heater tempearture for the CNT tip (d) making the new CNT probe more sensitive to sample temperature changes.

The immediate obvious qualitative result, evident in Figure 8 (a), is that the CNT tip acts as a heat distributor that cools down a significant volume of the SP tip including the Pd resistive heater. For comparison, in the original SP setup (Figure 8 (b)) only the very end of the tip apex is cooled. This means that the SP-CNT tip is closer in temperature to that of the sample and would provide a better reading of the sample temperature than a normal SP tip. In our models we have evaluated CNT's of different diameters. CNT's of diameters of about 10-20 nm showed significant temperature drop along the nanotube itself, whereas thicker multiwall tubes of about 50 nm diameter showed a significant heat transfer to the probe and will be more desirable for thermal probes. Possible sharpening of 50 nm probes can further improve their spatial resolution.

The quantitative analysis of the new CNT probe (Fig. 8 (c), (d)) shows that the new probe increases the thermal response of the SP SThM probe by approximately one order of magnitude with the sensitivity expanded well above to 100 $Wm^{-1}K^{-1}$. It should be noted that due to high anisotropy of thermal conductivity in CNT, the heat exchange between the SP probe and the CNT will be mainly diffusive, whereas the heat conduction along its direction $L \sim 1$ μm is, at least partially ballistic, given the MFP of phonons in CNT at room temperature $\Lambda_{CNT} \sim 700$ nm[36] that may lead to some reduction of performance of such probe compared to evaluated in Fig 8. Performance of the new probe should be appropriate for the exploration of Si nanostructures with Cu, Au, Al components and a significant step toward exploration of highly conductive materials like graphene and diamond.

Given the non-negligible thermal response of the SThM-CNT probe for higher thermal conductivity materials, such as metals and crystalline semiconductors, as we mentioned earlier, the tip-surface interfacial contact resistance could play a more significant role in these probes. Also, the thermal resistance between the CNT and the SP probe, and anisotropy of the thermal conductivity of CNT that may reach one to two orders of magnitude[37] might also change the characteristics of these probes.

First, we found that the thermal interface resistance between the side of the CNT and the SThM probe, as well as anisotropy of CNT thermal conductivity would make a small difference on the order of 10 to 20%, mainly due to the extended length of CNTs contact with the side of SP tip allowing efficient heat exchange.

At the same time, by introducing the interface thermal resistance similarly to the approach used in section B, we find that for low conductivity of the samples of 1-10 $Wm^{-1}K^{-1}$, in case of SThM-CNT probe the interface resistance may reduce its sensitivity by 10-30%. This drop may be even stronger for high thermal conductivity sample such as Si, with the thermal sensitivity may be reduced by 40% to 60% depending on the quality of the thermal contact. While this is not negligible, we believe that the ability to measure such high thermal conductivity materials with the nanoscale resolution is still very important achievement in measurement science.

## IV. Conclusions

We have built quantitative physical models for two major modalities of SThM nanoprobes and validated these models using experimental measurements of the thermal response of probes in air and vacuum, as well as directly in the SPM operation in nanoscale contact with the sample. These models provide an essential basis for analysing SThM performance in the exploration of nanostructures. In particular, using these models we have found that for the high thermal conductivity materials such as Si and metals, the probe apex and surface contact limits the heat transfer to the sample and the probe. This therefore hinders the SThM ability to explore nanostructures made of such materials. Our analysis shows that the best thermal nanoprobe would have a thermal sensor positioned near the apex of the probe, and a high thermal conductivity material in contact with the sample. We have explored the thermal performance of such SThM probe that would be based on the existing silicon oxide (or silicon nitride) probe with a Pd resistive element with high thermal conductivity 50 nm diameter multiwall CNT at the end of the probe. We showed that this new probe will have a significantly superior thermal response which exceeds that of existing SThM probes by an order of magnitude. This would expand SThM ability to explore high thermal conductivity materials, widely used in semiconductor industry and nanoelectronic devides, such as Si, Cu, Au, Al as well graphene and CNT based nanostructures, with nanoscale resolution.

## Acknowledgements

Authors acknowledge input of Craig Prater and Roshan Shetty from Anasys Instruments for insight, scientific discussions and support related to the SThM development. We are also grateful to Hubert Pollock, Olaug Grude, Bob Jones and late Azeddine Hammiche for insight in using SThM, Maria Timofeeva for analysis of thermal transport in anisotropic materials, Mark Rosamond for SEM analysis of the SP probes. OVK acknowledges support from the EPSRC grants EP/G015570/1, EPSRC-NSF grant EP/G06556X/1 and EU FP7 GRENADA and FUNPROBE grants.